\documentclass[12pt,draftclsnofoot,onecolumn]{IEEEtran}
\usepackage{amssymb}
\usepackage{amsmath}
\usepackage{amsfonts}
\usepackage{graphicx}
\usepackage{subfigure}
\usepackage{cite}
\usepackage{enumerate}
\usepackage{url}

\begin{document}

\title{\LARGE{Sub-channel Assignment, Power Allocation and User Scheduling for Non-Orthogonal Multiple Access Networks}}
\author{
\IEEEauthorblockN{
\normalsize{Boya Di}\IEEEauthorrefmark{1},
\normalsize{Lingyang Song}\IEEEauthorrefmark{1},
\normalsize{and Yonghui Li}\IEEEauthorrefmark{2}\\}
\IEEEauthorblockA{
\IEEEauthorrefmark{1}\normalsize{School of Electrical Engineering and Computer Science, Peking University, China.} \\
\IEEEauthorrefmark{2}School of Electrical and Information Engineering, The University of Sydney, Australia.}\\
%\IEEEauthorrefmark{2}\normalsize{Electrical and Computer Engineering Department, University of Houston, Houston, TX, USA.} \\
%\IEEEauthorrefmark{3}\normalsize{ZTE Corporation, Shenzhen, Guangdong, China.} \\ }
\thanks{Part of the material in this paper was presented in IEEE Globecom, San Diego, CA, Dec.~2015~\cite{BSYL-2015}.}
}

\maketitle

\begin{abstract}

In this paper, we study the resource allocation and user scheduling problem for a downlink non-orthogonal multiple access network where the base station allocates spectrum and power resources to a set of users. We aim to jointly optimize the sub-channel assignment and power allocation to maximize the weighted total sum-rate while taking into account user fairness. We formulate the sub-channel allocation problem as equivalent to a many-to-many two-sided user-subchannel matching game in which the set of users and sub-channels are considered as two sets of players pursuing their own interests. We then propose a matching algorithm which converges to a two-side exchange stable matching after a limited number of iterations. A joint solution is thus provided to solve the sub-channel assignment and power allocation problems iteratively. Simulation results show that the proposed algorithm greatly outperforms the orthogonal multiple access scheme and a previous non-orthogonal multiple access scheme.
\end{abstract}

\begin{IEEEkeywords}
Non-orthogonal multiple access, resource allocation, scheduling problem, matching game.
\end{IEEEkeywords}

\newpage

%%%%%%%%%%%%%%%%%%%%%%%
\section{Introduction}%
%%%%%%%%%%%%%%%%%%%%%%%
Orthogonal frequency division multiple access (OFDMA), as one of the prominent multicarrier transmission techniques, has been widely adopted in the 4th generation (4G) mobile communication systems such as LTE and LTE-Advanced~\cite{ARBNT-2010} to combat narrow-band interference. Multiple users are allocated orthogonal resources in frequency domain in order to achieve multiplexing gain with reasonable complexity~\cite{LBYSCZ-2015}. However, due to the explosive growth of data traffic in mobile Internet, there are increasing demands for high spectrum efficiency and massive connectivity in the 5th generation (5G) wireless communications~\cite{JXHRHGS-2014}. To address these challenges, various new multiple access techniques have been recently proposed such as Interleave Division Multiple Access (IDMA)~\cite{LLKW-2006}, Low Density Spreading (LDS)~\cite{MMR-2012}, and Non-Orthogonal Multiple Access (NOMA)~\cite{KH-2012}. Among these techniques, NOMA has a lower receiver complexity and achieves significant improvement in spectral efficiency and massive connectivity by allowing multiple users to share the same sub-channel in power domain, and thus, it has been considered as a promising candidate for future access technologies~\cite{LBYSCZ-2015}.

%Several technical challenges are studied in the open literature for the NOMA scheme.
Unlike the OFDMA scheme in which one sub-channel can only be assigned to one user, multiple users can share the same sub-channel simultaneously in the NOMA scheme, creating the inter-user interference over each sub-channel. To tackle this problem, various multi-user detection~(MUD) techniques such as the successive interference cancellation (SIC)~\cite{JJP-2014} can be applied at the end-user receivers to decode the received signals. Through power domain multiplexing at the transmitter and SIC at the receivers, NOMA can achieve a capacity region which significantly outperforms the orthogonal multiple access (OMA) schemes~\cite{DP-2005}. Recently different aspects of the NOMA schemes have been discussed in several works~\cite{YAYT-2013}$-$\hspace{-0.01cm}\cite{YYATAK-2013}. In~\cite{YAYT-2013}, the concept of basic NOMA with SIC was introduced and its performance was compared with the traditional OFDMA scheme through a system-level evaluation. A low-complexity power allocation method for NOMA with SIC receiver was discussed in~\cite{AAH-2014} by exploiting a tree search algorithm. In~\cite{ZZPH-2014}, the ergodic sum-rate and outage probability were derived with fixed power allocation. In~\cite{YDZR-2016}, the authors studied the subcarrier and power allocation problem in the NOMA system. They assumed that two users can share the same sub-channel simultaneously, and an optimal solution was approximated via the monotonic optimization approach. In~\cite{YYATAK-2013}, the authors discussed the user fairness in an uplink NOMA scheme for the wireless network in which the ML-MUD was applied and a link-level performance was evaluated.

However, so far few works have considered the joint sub-channel and power allocation problem for a general NOMA system. In most existing works~\cite{YAYT-2013}$-$\hspace{-0.01cm}\cite{YYATAK-2013}, either the power allocation is fixed~\cite{ZZPH-2014}, or the sub-channel allocation schemes are performed in random or greedy methods~\cite{AAH-2014}. In this paper we consider a downlink NOMA wireless network in which the base station (BS) assigns the sub-channels to a set of users and allocates different levels of power to them. Each user has access to multiple sub-channels and each sub-channel can be shared by multiple users. For the users sharing the same sub-channel, SIC is adopted at the receiver to remove the inter-user interference. Note that the sub-channel and power allocation are closely coupled with each other, influencing the system spectral efficiency together. We then formulate the joint sub-channel and power allocations as a non-convex weighted total sum-rate maximization problem in which user fairness is considered. This is an NP-hard problem and remains as an open problem in the literature due to its combinatorial nature and co-channel interference.

To tackle the above problem, we decouple the sub-channel and power allocation problems, and propose a joint solution in which the sub-channel and power allocation are solved iteratively. Aiming at finding an effective algorithm, we recognize that the sub-channel allocation problem can be regarded as a matching process with externalities. The users and sub-channels can be considered as two sets of players to be matched with each other to achieve the maximum weighted sum-rate, while interdependencies exist among the users due to the inter-user interference. We thus solve this problem by utilizing the matching games~\cite{7AM-1992,ECABA-2011}, which provide an adaptive and low-complexity framework to solve the resource allocation problem with combinatorial nature~\cite{22BLHVL-2013, BSYL-2014,23SSVBM-2014}. The sub-channel allocation problem is then formulated as a many-to-many two-sided matching problem with externalities, which is more complex than traditional two-sided matching problems without externality. Two novel user-subchannel swap-matching algorithms~(USMA) are developed in which a stable matching and a global optimal matching can be reached, respectively.

The main contributions of this paper can be summarized as follows. We formulate a joint sub-channel and power allocation problem for a downlink NOMA network to maximize the weighted total sum-rate. To tackle this NP-hard joint optimization problem, we decouple the sub-channel and power allocation problems as a many-to-many matching game with externalities and a geometric programming, respectively. For the matching game, we propose two matching algorithms (USMA-1 and USMA-2) in which a two-sided exchange-stable matching is formed after a small number of iterations in USMA-1. With a sufficiently large number of iterations in USMA-2, a global optimal matching can be obtained, along with the power allocation scheme approaching the joint optimal solution. We analyze the proposed matching algorithms in terms of the stability, convergence, complexity, and optimality. Simulation results show that our proposed algorithms can achieve a better performance than a previous resource allocation scheme proposed in~\cite{AAYYAT-2013}, a random allocation scheme and the OFDMA scheme.

Note that in our conference version~\cite{BSYL-2015}, we only considered the basic sum rate utility and applied an extended GS algorithm which cannot fully depict the externalities caused by the co-channel interference. Compared to our previous work~\cite{BSYL-2015}, we significantly extend the conference version in several major aspects. First, we consider a more general weighted sum-rate as the major optimization metric in this paper, and prove that it is an NP-hard problem. To tackle this challenging problem and fully explore the impact brought by co-channel interference, we then propose a novel swap-matching algorithm, and some new theories are developed, such as the swap-blocking pair, the stability and the swap operation. Furthermore, the user fairness is considered in the simulation study.

The rest of this paper is organized as follows. In Section \uppercase\expandafter{\romannumeral2}, we describe the system model. In Section \uppercase\expandafter{\romannumeral3}, we formulate the resource allocation as a weighted sum-rate maximization problem, and the sub-channel and power allocation problems are decoupled as a many-to-many matching game with externalities and a geometric programming, respectively. The proposed algorithms and related properties are analyzed in Section \uppercase\expandafter{\romannumeral4}. Simulation results are presented in Section \uppercase\expandafter{\romannumeral5}. Finally, we conclude the paper in Section \uppercase\expandafter{\romannumeral6}.

%\vspace{-0.2cm}
%%%%%%%%%%%%%%%%%%%%%%%%%%%%%%%%%%%%%%%%%%%%%%%
\section{System Model}%
\begin{figure}[!t]
\centering
\includegraphics[width=4.5in]{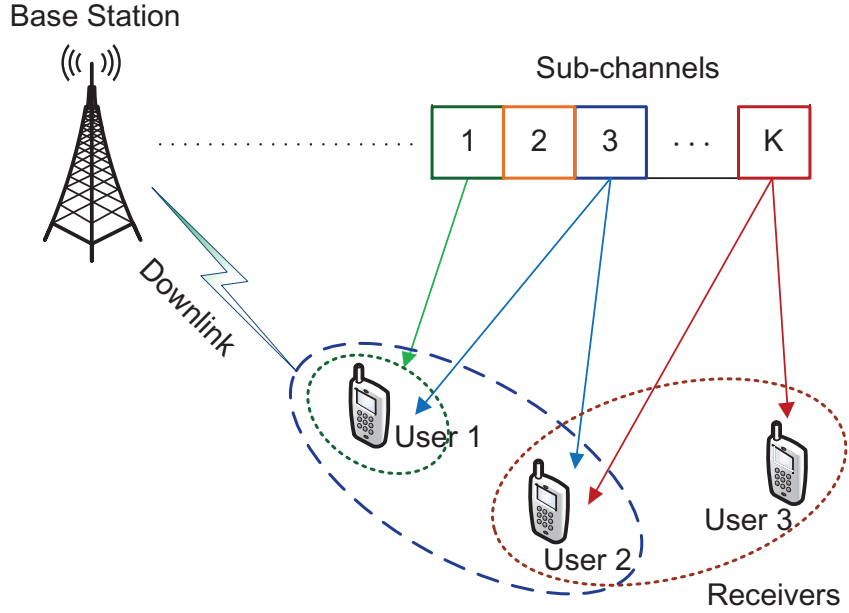}
\caption{System model of the NOMA networks.} \label{system_model}
\end{figure}

%%%%%%%%%%%%%%%%%%%%%%%%%%%%%%%%%%%%%%%%%%%%%%%
Consider a downlink single-cell NOMA network as shown in Fig.~\ref{system_model}, in which a single BS transmits the signals to a set of mobile users\footnote{Here we assume that each user has a fixed position. In a low-mobility case, if the channels do not change significantly within one time slot, the resource allocation scheme discussed in this paper can still work well. For a high-mobility case, the performance of users in NOMA is likely to degrade due to inaccurate channel estimation and inevitable frequency offset.} denoted by ${\cal{M}} = \left\{ {1, \cdots ,M} \right\}$. The BS divides the available bandwidth to a set of sub-channels, denoted by ${\cal{K}} = \left\{ {1, \cdots ,K} \right\}$. We assume that the BS has the full knowledge of the channel side information (CSI)\footnote{According to 3GPP TS 36.213~\cite{3GPP36}, the BS broadcasts training symbols to all the mobile stations (MSs). The MSs estimate the downlink channel and feed back the CSI to the BS through uplink feedback channels. Based on CSI, the BS then allocates the subcarriers and different power levels to the users.}. Based on the CSI of each channel, the BS assigns a subset of non-overlapping sub-channels to the users and allocates different levels of power to the users. According to the NOMA protocol~\cite{KH-2012}, one sub-channel can be allocated to multiple users, and one user can receive from the BS through multiple sub-channels. The power allocated to user $M_j \in \cal{M}$ over sub-channel ${SC}_k$ is denoted by $p_{k, j}$, satisfying $\sum\nolimits_{{\rm{k}} \in \mathcal{K}} {\sum\nolimits_{j \in \mathcal{M}} {{p_{k,j}}} }  \le {P_s}$ where $P_s$ is the total transmitted power of the BS. We consider a block fading channel, for which the channel remains constant within a time-slot, but varies independently from one to another. The complex coefficient of ${SC}_k$ between user $M_j$ and the BS is denoted by ${h_{k,j}} = {g_{k,j}}/\mathcal{D}\left( {{d_j}} \right)$, where $g_{k,j}$ denotes the Rayleigh fading channel gain, $d_j$ is the distance between user $M_j$ and the BS, and $\mathcal{D}\left(  \cdot  \right)$ is the path loss function. Let ${\cal{S}}_k$ be the set of active users over sub-channel ${SC}_k$, and $x_{k,i}$ be the transmitted symbol of user $M_i$ over sub-channel ${SC}_k$. The signal that user $M_j$ receives over sub-channel ${SC}_k$ is then given by
\begin{equation} \label{receive_signal}
{y_{k,j}} = {h_{k,j}}\sum\limits_{i \in {{\cal{S}}_k}} {\sqrt {{p_{k,i}}} {x_{k,i}}}  + {n_{k,j}},
\end{equation}
where ${n_{k,j}} \sim \mathcal{CN}\left( {0,{\sigma _n}^2} \right)$ is the additive white Gaussian noise~(AWGN) for user $M_j$ over ${SC}_k$, and ${\sigma _n}^2$ is the noise variance.

%\footnote{For the imperfect case in which there exists error propagation during the SIC decoding, the HARQ process is then required in this case to guarantee the NOMA performance~\cite{AAYYAT-2013}. For convenience, we assume successful SIC decoding in this paper.}

%Here we assume that each user has a fixed position. For a high-mobility case, the performance of users in NOMA is likely to degrade due to inaccurate channel estimation and inevitable frequency offset.

Since sub-channel ${SC}_k$ can be utilized by a subset of users, ${\cal{S}}_k$, the signal of any user ${M_j} \in \mathcal{S}_k$ causes interference to other user ${M_{j'}} \in {\mathcal{S}_k}$. To demodulate the target message, each user $M_j$ adopts SIC\footnote{As a non-linear multi-user receiver, SIC can achieve better performance than traditional linear receivers such as LMMSE with an affordable complexity increase. In addition, it also has a much lower complexity compared to the optimal ML detector, which makes the problem more tractable in the NOMA system. Specifically, SIC can significantly reduce the receiver complexity from exponential complexity in optimal maximum likelihood (ML) detection, i.e., $O\left( {{{\left| \mathbb{X} \right|}^{{d_f}}}} \right)$, to polynomial complexity $\mathcal{O}\left( {{d_f}^3} \right)$, where $\left| \mathbb{X} \right|$ denotes the cardinality of the constellation set $\mathbb{X}$.} after receiving the superposed signals~\cite{JJP-2014}. In general, the users with higher channel gains are allocated low power levels and their signals can be recovered after all users with higher power levels are recovered in the SIC decoding, while the users with lower channel gains have large power assignment levels and their signals are recovered by treating the users' signals with lower power levels as the noise in the SIC decoding~\cite{JJP-2014}-\cite{AAH-2014}. Thus, the optimal order of SIC decoding is in the order of the increasing channel gains normalized by the noise. To be specific, the receiver of user $M_j \in {\cal{S}}_k$ can cancel the interference from any other user $M_i$ in ${\cal{S}}_k$ with channel gain ${\left| {{h_{k,i}}} \right|^2}/{n_{k,i}} < {\left| {{h_{k,j}}} \right|^2}/{n_{k,j}}$, i.e., user $M_j$ first decodes the signal from user $M_i$, then it subtracts this signal and decodes its target signal $x_{k, j}$ correctly from received signal $y_{k,j}$. For those users with higher channel gain than user $M_j$'s, $M_j$ regards their signals as noise and decodes $x_{k,j}$. The decoding order described above guarantees that the upper bound on the capacity region can be reached~\cite{B-1974,WJ-2004}, i.e., the capacity of user $M_j$ over sub-channel ${SC}_k$ within one time slot is given by
\begin{equation} \label{throughput_single}
{R_{k,j}} = {\log _2}\left( {1 + \frac{{{p_{k,j}}{{\left| {{h_{k,j}}} \right|}^2}}}{{{n_{k,j}} + {I_{k,j}}}}} \right),
\end{equation}
where $I_{k,j}$ is the interference that user $M_j$ receives from other users in $\mathcal{S}_k$ over sub-channel ${SC}_k$,
\begin{equation} \label{interference_user}
{I_{k,j}} = \sum\nolimits_{i \in \left\{ {{{\cal{S}}_k}|\frac{{{{\left| {{h_{k,i}}} \right|}^2}}}{{{n_{k,i}}}} > \frac{{{{\left| {{h_{k,j}}} \right|}^2}}}{{{n_{k,j}}}}} \right\}} {{p_{k,i}}{{\left| {{h_{k,j}}} \right|}^2}}.
\end{equation}
We assume that user $M_j$ can decode the signals from user $M_i$ correctly if ${\left| {{h_{k,i}}} \right|^2}/{n_{k,i}} < {\left| {{h_{k,j}}} \right|^2}/{n_{k,j}}$, since $R_{k,j}^i \ge {R_{k,i}}$, in which $R_{k,j}^i$ denotes the rate for user $M_j$ to detect user $M_i$'s signals, i.e., $R_{k,j}^i = {\log _2}\left( {1 + \frac{{{p_{k,i}}{{\left| {{h_{k,j}}} \right|}^2}}}{{{n_{k,j}} + {{\left| {{h_{k,j}}} \right|}^2}\sum\nolimits_{m \in {S_i}} {{p_{k,m}}} }}} \right)$. Note that SIC performed at the user receiver may cause considerable complexity ${\rm \mathcal{O}}\left( \left| {{\mathcal{S}_{k}}} \right|^3 \right)$~\cite{LBYSCZ-2015}. Therefore, considering the complexity caused by decoding, we assume that at most $d_f$ users can share one sub-channel at the same time, i.e., $\left| {{\mathcal{S}_k}} \right| \le {d_f}$. When ${d_f} \ll M$, the decoding complexity at the receiver is much reduced to a tolerable level. Naturally, the achievable rates of ${SC}_k$ can be expressed as
\begin{equation} \label{rate_subchannel}
{R_{S{C_k}}} = \sum\limits_{j \in {\mathcal{S}_k}} {{w_j}} {\log _2}\left( {1 + \frac{{{p_{k,j}}{{\left| {{h_{k,j}}} \right|}^2}}}{{{n_{k,j}} + {I_{k,j}}}}} \right)
\end{equation}

%To better describe the scheduling policy of the users, i.e., the allocation of the sub-channels to the set of users, we introduce a $K \times M$ sub-channel matrix $\Gamma$ in which the binary element ${\gamma _{k,j}}\left( t \right)$ denotes whether sub-channel ${SC}_k$ is allocated to user $M_j$ in time slot $t$.

\section{Problem Formulation}
In this section, sub-channel and power allocation for the downlink NOMA network will be discussed. We consider the user fairness and formulate the sub-channel and power allocation problem as a weighted sum-rate maximization problem, and then model the sub-channel assignment problem and power allocation problem as a two-sided many-to-many matching game with externalities and a geometric programming (GP), respectively~\footnote{It is worth noting that the matching theory along with GP is just one efficient method to solve the resource allocation problem. Other feasible methods such as the optimization theory~\cite{L-2015} and the coalition formation games~\cite{23CEW-2012,BTLH-2016} may be considered as well.}.

\subsection{Weighted Sum-rate Maximization Problem Formulation}
We consider the user fairness in terms of capacity, and introduce a weight factor $w_j$ for each user $M_j$ to adjust its priority when allocating resources. Based on the proportional fairness scheduler~\cite{K-1997,JZWZ-2004}, we assume that $w_j$ is inversely proportional to the average rate of user $M_j$ in previous time slots. We then introduce a $K \times M$ sub-channel matrix $\emph{\textbf{B}}$ in which the binary element ${b _{k,j}}$ denotes whether sub-channel ${SC}_k$ is allocated to user $M_j$. The system performance can be evaluated by the total sum-rate of all users, also known as the total utility of the system, which is shown as below:
\begin{equation}
{U_{total}} = \sum\limits_{j \in \mathcal{M}} {{w_j}} \sum\limits_{k \in \mathcal{K}} {{b_{k,j}}} {\log _2}\left( {1 + \frac{{{p_{k,j}}{{\left| {{h_{k,j}}} \right|}^2}}}{{{n_{k,j}} + {I_{k,j}}}}} \right).
\end{equation}
Our objective is to maximize the total sum-rate of the system by setting the variables $\left\{ {{p_{k,j}},{b _{k,j}}} \right\}$. The optimization problem is then formulated as:
\begin{subequations} \label{system_optimization}
\begin{align}
\mathop {\max }\limits_{{b _{k,j}},{p_{k,j}}} & {U_{total}} \label{system_41}\\
\textbf{\emph{s.t.:}} & \sum\limits_{j \in \cal{M}} {{b _{k,j}}}  \le {d_f},\forall k \in {\cal{K}},\label{system_5}\\
& \sum\limits_{k \in \mathcal{K}} {{b_{k,j}}}  \le {d_v},\forall j \in \mathcal{M}, \label{system_7}\\
& {b _{k,j}} \in \left\{ {0,1} \right\},\forall k \in {\cal{K}},j \in {\cal{M}},\label{system_2}\\
& \sum\limits_{k \in \mathcal{K}} {\sum\limits_{j \in \mathcal{M}} {{p_{k,j}}}  \le {P_s}},\label{system_11}\\
& {p_{k,j}} \ge 0,\forall k \in {\cal{K}},j \in \cal{M}.\label{system_31}
\end{align}
\end{subequations}
Constraints $\left( \ref{system_5} \right)$-$\left( \ref{system_2} \right)$ ensure that each sub-channel can only be assigned to at most $d_f$ users, and each user can only occupy at most $d_v$ sub-channels in terms of user fairness. Due to the limited transmitted power of the BS, power variables must satisfy constraints $\left( \ref{system_11} \right)$ and $\left( \ref{system_31} \right)$.

Note that this is a non-convex optimization problem due to the binary constraint in $\left( \ref{system_2} \right)$ and the existence of the interference term in the objective function~\cite{LP-2005}. Since the channel coefficients vary in different channels due to frequency selectivity, the noise power is nonconstant if normalized, which makes it difficult to convert the problem into a convex one~\cite{LZ-2008}. In Proposition 1, we prove that the non-convex optimization problem $\left( \ref{system_optimization} \right)$ is also an NP-hard problem.

\textbf{Proposition 1:} The weighted sum-rate maximization problem in $\left( \ref{system_optimization} \right)$ is NP-hard.

\begin{proof}
See Appendix A.
\end{proof}

A tractable version of problem $\left( \ref{system_optimization} \right)$ can be presented through the following proposition, which evaluates how the constraints influence the solution of this problem.

\textbf{Proposition 2:} If we remove constraint (6c) and rewrite the power constraint in (6e) as $\sum\nolimits_{j \in \mathcal{M}} {{{\rm{p}}_{k,j}}}  \le {P_k} = {P_s}/K,\forall k \in \mathcal{K}$, then the variant of the resource allocation problem with equal weight factor is easy to handle with a closed-form optimal solution.

\begin{proof}
See Appendix B.
\end{proof}

%As shown in~\cite{AAH-2014}, NOMA is superior to OMA in terms of both the total sum rate and user fairness from the prospective of capacity region.

As observed in the objective function $\left( \ref{system_41} \right)$, multi-user power allocation and sub-channel allocation are coupled with each other in terms of the total sum-rate. Considering the computational complexity, we decouple these two subproblems, and propose an iterative algorithm in which the power and sub-channel allocations are performed in an iterative way to obtain a joint solution for problem $\left( \ref{system_optimization} \right)$. Given the sub-channel allocation, the power allocation problem can then be formulated as geometric programming (GP)~\cite{SS-2007,S-2008}, i.e., a convex problem in which the global-optimal solution can be obtained via efficient interior point methods~\cite{SL-2004}. Given the power allocation, sub-channel allocation can be formulated as a many-to-many two-sided matching problem, which can be solved by utilizing the matching games.

\subsection{Geometric Programming for Power Allocation}
Suppose sub-channel allocation is settled, i.e., sub-channel matrix $\emph{\textbf{B}}$ and the set of $\mathcal{S}_k$ for each ${SC}_k$ are given. Problem $\left( \ref{system_optimization} \right)$ can then be rewritten as
\begin{subequations} \label{power_allocation}
\begin{align}
\mathop {\max }\limits_{{p_{k,j}}} {\rm{ }} & \sum\limits_{k \in \mathcal{K}} {\sum\limits_{j \in {\mathcal{S}_k}} {{w_j}} {R_{k,j}}} \label{system_4}\\
\textbf{\emph{s.t.:}} & \sum\limits_{k \in \mathcal{K}} {\sum\limits_{j \in {\mathcal{S}_k}} {{p_{k,j}}}  \le {P_s}},\label{system_1}\\
& {p_{k,j}} \ge 0,\forall k \in {\cal{K}},j \in {\mathcal{S}_k}.\label{system_3}
\end{align}
\end{subequations}
We can reformulate this problem as geometric programming by utilizing the method shown in~\cite{S-2008}. The users in $\mathcal{S}_k$ can be resorted according to the channel gains in a decreasing order. Thus, by setting ${m_{k,{\pi _k}\left( j \right)}} = {n_{{\pi _k}\left( j \right)}}/{\left| {{h_{k,{\pi _k}\left( j \right)}}} \right|^2}$, we have ${m_{k,{\pi _k}\left( 1 \right)}} < {m_{k,{\pi _k}\left( 2 \right)}} <  \cdots  < {m_{k,{\pi _k}\left( {{d_f}} \right)}}$, in which ${{\pi _k}\left( \cdot \right)}$ is in order of decreasing channel SNRs over sub-channel ${SC}_k$.

\textbf{Proposition 3:} With SIC for decoding at the receivers, problem $\left( \ref{power_allocation} \right)$ can be converted into the following GP with $\left\{ {{R_{k,j}}} \right\}$ as variables:
\begin{subequations} \label{power_GP}
\begin{align}
\mathop {\min }\limits_{{R_{k,j}}} & \log {e^{ - \sum\limits_{k \in K} {\sum\limits_{j \in {S_k}} {{w_j}{R_{k,j}}} } }} \label{system_4}\\
\textbf{\emph{s.t.:}} &\log {e^{ - {R_{k,j}}}} \le 0,\forall k \in {\cal K},j \in {{\cal S}_k},\label{system_1}\\
& \log \sum\limits_{k \in {\cal K}} {\sum\limits_{j \in {{\cal S}_k}} {\left( {\frac{{{m_{k,{\pi _k}\left( j \right)}} - {m_{k,{\pi _k}\left( {j - 1} \right)}}}}{{{P_s} + \sum\nolimits_{t = 1}^K {{m_{t,{\pi _t}\left( {{d_f}} \right)}}} }}} \right)} }  \times {e^{\sum\nolimits_{i = j}^{{d_f}} {{R_{k,{\pi _k}\left( i \right)}}} \ln 2}} \le 0.\label{system_3}
\end{align}
\end{subequations}
\begin{proof}
See Appendix C.
\end{proof}
The objective function of problem $\left( \ref{power_GP} \right)$ is linear, and the constraints are convex~\cite{SL-2004}, which is in consistent to the form of GP~\cite{SS-2007}. Therefore, problem $\left( \ref{power_GP} \right)$ can be solved by utilizing interior point methods~\cite{SL-2004}.

\subsection{Many-to-many Two-sided Matching Game Formulation for Sub-channel Allocation}

\subsubsection{Definition}

To develop a low-complexity sub-channel allocation algorithm, we consider the set of users, $\cal{M}$, and the set of sub-channels, $\cal{K}$, as two disjoint sets of selfish and rational players aiming to maximize their own benefits\footnote{Note that it is still the BS that determines the sub-channel allocation. The BS can obtain a sub-channel allocation scheme by performing such an algorithm in which it considers the set of sub-channels and users as selfish and rational players.}. Each player can exchange information with one another without extra signaling cost, since the CSI is known to the BS, i.e., the players have complete information about each other. Specifically, if sub-channel ${SC}_k$ is assigned to user $M_j$, then we say $M_j$ and ${SC}_k$ are matched with each other and form a \emph{matching pair}. A matching is defined as an assignment of sub-channels in $\mathcal{K}$ to users in $\cal{M}$, formally presented as

\textbf{Definition 1:} Given two disjoint sets, ${\cal{M}} = \{ 1,2, \cdots, M\}$ of the users, and ${{\cal{K}}} = \{ 1,2, \cdots, K\}$ of the sub-channels, a many-to-many \emph{matching} $\Psi$ is a mapping from the set $\cal{M} \cup {{{\mathcal{K}}}} \cup$$ \left\{ 0 \right\}$ into the set of all subsets of $\cal{M} \cup {\cal{K}} \cup$$ \left\{ 0 \right\}$ such that for every $M_j \in \cal{M}$, and ${SC}_k \in {{\cal{K}}}$:

1) $\Psi \left( {{M_j}} \right) \subseteq {{{\cal{K}}}}$;

2) $\Psi \left( {S{C_k}} \right) \subseteq {\cal{M}}$;

3) $\left| {\Psi \left( {S{C_k}} \right)} \right| \le {d_f}$;

4) $\left| {\Psi \left( {{M_j}} \right)} \right| \le {d_v}$;

5) $S{C_k} \in \Psi \left( {{M_j}} \right) \Leftrightarrow {M_j} \in \Psi \left( {S{C_k}} \right)$.

Condition 1) states that each user is matched with a subset of sub-channels, and condition 2) implies that each sub-channel is matched with a subset of users. Taking into account the tolerable complexity of the decoding technique at the receiver, we set the size of $\Psi \left( {S{C_k}} \right)$ no larger than $d_f$ and that of $\Psi \left( {{M_j}} \right)$ no larger than $d_v$, as expressed in conditions 3) and 4).

\textbf{Remark 1}: The matching game formulated above is a many-to-many matching game with \emph{externalities}, also known as the peer effects.
\begin{proof}
See Appendix D.
\end{proof}

\subsubsection{Preference Lists of the players}

Influenced by the peer effects~\cite{ECABA-2011}, the outcome of this matching game greatly depends on the dynamic interactions between the users. To better describe the competition behavior and decision process of each player, we assume that each player has preferences over the players of the other set, and we introduce a preference relation $\succ$ for both users and sub-channels. Specifically, for any user $M_j \in {\cal{M}}$, its preference ${ \succ _{{M_j}}}$ over the set of sub-channels can be described as follows. For any two sub-channels $S{C_k},S{C_{k'}} \in {\mathcal{K}}$, $k \ne k'$, and any two matchings $\Psi ,\Psi '$, $S{C_k} \in \Psi \left( {{M_j}} \right)$, $S{C_{k'}} \in \Psi '\left( {{M_j}} \right)$:  %要说明一下用户的utility就是rate
\begin{equation} \label{preference_user}
\left( {S{C_k},\Psi } \right){ \succ _{{M_j}}}\left( {S{C_{k'}},\Psi '} \right) \Leftrightarrow {R_{kj}}\left( \Psi  \right) > {R_{k'j}}\left( {\Psi '} \right)
\end{equation}
indicates that user $M_j$ prefers $S{C_k}$ in $\Psi$ to $S{C_{k'}}$ in $\Psi '$ only if $M_j$ can achieve a higher rate over $S{C_k}$ than over $S{C_{k'}}$. We assume that $\Psi$ and $\Psi '$ are allowed to refer to the same matching. Similarly, for any sub-channel $S{C_k} \in {\mathcal{K}}$, its preference ${ \succ _{S{C_k}}}$ over the set of users can be described as follows. For any two subsets of users $T,T' \subseteq \mathcal{M}$, $T \ne T'$, and any two matchings $\Psi ,\Psi '$, $T = \Psi \left( {S{C_k}} \right)$, $T' = \Psi' \left( {S{C_k}} \right)$:
\begin{equation} \label{preference_sub-channel}
\left( {T,\Psi } \right){ \succ _{S{C_k}}}\left( {T',\Psi '} \right) \Leftrightarrow {R_{S{C_k}}}\left( \Psi  \right) > {R_{S{C_k}}}\left( {\Psi '} \right)
\end{equation}
implies that ${SC}_k$ prefers the set of users $T$ to $T'$ only when ${SC}_k$ can get a higher rate from $T$.

\textbf{Remark 2}: Different from traditional matchings, each sub-channel's preference does not satisfy \emph{substitutability} any more.
\begin{proof}
See Appendix E.
\end{proof}

%有关many-to-many的难度分析看一下算法那本书
Note that a many-to-many matching model with externalities is more complicated than the conventional two-sided matching models. Under traditional definition of \emph{stable matching}\footnote{Traditional stable matching refers to a matching in which no two players from opposite sets prefer each other to at least one of their current matches such that they form a new matching pair together for the sake of their interests, i.e., there does not exist blocking pairs~\cite{7AM-1992} in a stable matching.} such as that in~\cite{7AM-1992}, there is no guarantee that a stable matching exists even in many-to-one matchings. In fact, it is computationally hard to find the stable matching even if it does exist~\cite{ECABA-2011}. Due to the lack of substitutability, traditional deferred acceptance algorithm~\cite{7AM-1992} and standard form of fixed point methods~\cite{FJ-2004} do not apply any more. Therefore, to solve this matching problem, we introduce the notion of \emph{switch matching}~\cite{ECABA-2011} and propose two matching algorithms in Section~\uppercase\expandafter{\romannumeral4}.

\section{Many-to-Many Matching Algorithm for NOMA}
Inspired by the many-to-one housing assignment problem with externalities~\cite{ECABA-2011}, we introduce the notions of \emph{switch matching} and \emph{two-sided exchange stability} into our many-to-many matching model, and propose two matching algorithms for the sub-channel allocation problem. %Note that the matching algorithm is performed by the BS without extra signaling cost compared to the traditional centralized methods.

\subsection{Design of Many-to-Many Matching Algorithm}

Different from the traditional deferred acceptance approach~\cite{7AM-1992}, the swapping behaviour of the users is considered in which every two users are arranged by the BS to exchange their matches while keeping other players' assignment the same.

To better depict how the interdependency of players' preference relation, i.e., peer effects, influences the matching game, we first introduce the concept of \emph{swap-matching} and \emph{swap-blocking pair} as below.

\textbf{Definition 2:} Given a matching $\Psi$ with $S{C_p} \in \Psi \left( {{M_i}} \right)$, $S{C_q} \in \Psi \left( {{M_j}} \right)$, and $S{C_p} \notin \Psi \left( {{M_j}} \right)$, $S{C_q} \notin \Psi \left( {{M_i}} \right)$, a \emph{swap matching} $\Psi _{jq}^{ip} = \Psi \backslash \left\{ {\left( {{M_i},S{C_p}} \right),\left( {{M_j},S{C_q}} \right)} \right\} \cup \left\{ {\left( {{M_i},S{C_q}} \right),\left( {{M_j},S{C_p}} \right)} \right\}$ is defined by the function $S{C_q} \in \Psi _{jq}^{ip}\left( {{M_i}} \right),S{C_p} \in \Psi _{jq}^{ip}\left( {{M_j}} \right)$ and $S{C_q} \notin \Psi _{jq}^{ip}\left( {{M_j}} \right),S{C_p} \notin \Psi _{jq}^{ip}\left( {{M_i}} \right)$.

To be more specific, a \emph{swap-matching} is a matching generated via a \emph{swap operation}\footnote{The swap operation is a two-sided version of the ``exchange" considered in~\cite{ECABA-2011,KD-2002}.} in which two players in the same set exchange their matches in the opposite set while keeping all other players' assignment the same. Note that the existence of swap operation is reasonable based on the fact that every two users can exchange information with each other in our matching model~\footnote{Since the BS performs the matching algorithm to determine the sub-channel allocation based on the CSI, it makes sense to assume that the users can exchange information with each other.}. One of the users involved in a swap-matching is allowed to be unmatched, thus allowing for unscheduled users to be active.

However, considering their own interests, the players involved in a swap operation may not be approved by each other. By introducing the concept of \emph{swap-blocking pair}, we evaluate the conditions under which the swap operations will be approved.

\textbf{Definition 3:} Given a matching $\Psi$ and a pair $\left( {{M_i},{M_j}} \right)$ with ${M_i}$ and ${M_j}$ matched in $\Psi$, if there exist $S{C_p} \in \Psi \left( {{M_i}} \right)$ and $S{C_q} \in \Psi \left( {{M_j}} \right)$ such that:

(i) $\forall t \in \left\{ {{M_i},{M_j},S{C_p},S{C_q}} \right\},\left( {\Psi _{jq}^{ip}\left( t \right),\Psi _{jq}^{ip}} \right){\ge_t}\left( {\Psi \left( t \right),\Psi } \right)$,

(ii) $\exists t \in \left\{ {{M_i},{M_j},S{C_p},S{C_q}} \right\},\left( {\Psi _{jq}^{ip}\left( t \right),\Psi _{jq}^{ip}} \right){ \succ _t}\left( {\Psi \left( t \right),\Psi } \right)$,
then swap matching $\Psi _{jq}^{ip}$ is \emph{approved}, and $\left( {{M_i},{M_j}} \right)$ is called a \emph{swap-blocking pair} in $\Psi$.

The definition implies that if a swap matching is approved, then the achievable rates of any player involved will not decrease, and at least one player's data rates will increase. Note that either the users or the sub-channels can initiate the swap, since their benefits are all directly related to the data rates.

\textbf{Corollary 1:} For a swap-matching $\Psi _{jq}^{ip}$, if ${p_i}{\left| {{h_{p,i}}} \right|^2} = \mathop {\min }\limits_{k \in {\mathcal{S}_p}} \left\{ {{p_k}{{\left| {{h_{p,k}}} \right|}^2}} \right\}$, ${p_j}{\left| {{h_{q,j}}} \right|^2} = \mathop {\min }\limits_{k \in {\mathcal{S}_q}} \left\{ {{p_k}{{\left| {{h_{q,k}}} \right|}^2}} \right\}$, ${p_i}{\left| {{h_{q,i}}} \right|^2} = \mathop {\min }\limits_{k \in {\mathcal{S}_q}\backslash j \cup \left\{ i \right\}} \left\{ {{p_k}{{\left| {{h_{q,k}}} \right|}^2}} \right\}$, and ${p_j}{\left| {{h_{q,j}}} \right|^2} = \mathop {\min }\limits_{k \in {\mathcal{S}_p}\backslash i \cup \left\{ j \right\}} \left\{ {{p_k}{{\left| {{h_{p,k}}} \right|}^2}} \right\}$, then as long as $M_i$ and $M_j$ propose to swap their matches with each other, $\Psi _{jq}^{ip}$ is approved.

\begin{proof}
See Appendix D.
\end{proof}

Based on the above definitions, we can then depict the users' behaviours in a matching with peer effects as below. Every two users can be arranged by the BS to form a potential swap blocking pair. The BS checks whether they can benefit each other by exchanging their matches without hurting the interests of corresponding sub-channels. Through multiple swap operations, we show how dynamic preferences of different players are associated with each other, and the matching games's externalities are well handled. The players keep executing approved swap operations so as to reach a stable status, also known as a two-sided exchange stable matching defined as below.

\textbf{Definition 4:} A matching $\Psi$ is \emph{two-sided exchange stable (2ES)} if it is not blocked by any swap-blocking pair $\left( M_i, M_j \right)$.

Note that the notion of stability we consider in this setting is similar to that of~\cite{FMWSM-2013} due to peer effects, but differs from the traditional one used in~\cite{7AM-1992}.

\subsection{Algorithm Description}

With the definition of stability, we introduce two user-subchannel matching algorithms (USMA-1 and USMA-2) to obtain a 2ES matching. These two algorithms are extended versions of the many-to-one matching algorithms proposed in~\cite{ECABA-2011}. Different from the many-to-one matchings, we consider the constraints $\left| {\Psi \left( {S{C_k}} \right)} \right| \le {d_f}$ and $\left| {\Psi \left( {{M_j}} \right)} \right| \le {d_v}$ in the USMA.

The key idea of USMA-1 is to keep considering approved swap matchings among the players so as to reach a 2ES matching. The algorithm is described in detail in Table \ref{alg-SM-1}, consisting of initialization phase and swap matching phase. In the initialization phase, a priority-based allocation scheme is applied. We assume that the larger a user's weight is, the higher priority it has when choosing its preferred set of available sub-channels. The swap matching phase contains multiple iterations in which the BS keeps searching for two users to form a swap-blocking pair, then they execute the swap matching if approved, and update the current matching. The iterations stop until no users can form new swap-blocking pairs and a final matching is determined.

\begin{table}[!t]
\renewcommand{\arraystretch}{1.0}
\caption{User-Subchannel Matching Algorithm (USMA-1)}
\label{alg-SM-1}
\centering
\begin{tabular}{p{150mm}}
\hline

\textbf{Step 1: \emph{Initialization Phase}}

    While there exist at least one user and one sub-channel are not fully matched~\footnote{A user $M_j$ is not fully matched if $\left| {\Psi \left( {{M_j}} \right)} \right| < {d_v}$, and ${SC}_k$ is not fully matched if $\left| {\Psi \left( {S{C_k}} \right)} \right| < {d_f}$.} simultaneously:
    \begin{enumerate}
        \item ${j^*} = \arg \mathop {\max }\limits_{i \in \mathcal{M}} \left\{ {{w_i}} \right\}$.
        \item User $M_{j^*}$ matches with its most preferred subset of sub-channels each of which is not fully matched.
        \item Remove $M_{j^*}$ from $\cal{M}$.
    \end{enumerate}

{\bf Step 2: \emph{Swap matching phase}.}

In each round, for every matched user $M_j \in \cal{M}$,
\begin{enumerate}
\item The BS searches $\mathcal{M}\backslash \left\{ {{M_j}} \right\}$ for a swap-blocking pair $\left( M_i, M_j \right)$ along with $S{C_p} \in \Psi \left( {{M_i}} \right)$ and $S{C_q} \in \Psi \left( {{M_j}} \right)$ such that $\Psi _{jq}^{ip}$ is never executed in current round; otherwise go to Step 2-.
\item If $\Psi _{jq}^{ip}$ is approved, $M_i$ exchanges its match ${SC}_p$ with $M_j$ for ${SC}_q$. Set $\Psi  = \Psi _{jq}^{ip}$.
\item Else, $M_j$ keeps its matches.
\item Go back to Step-2-1.
\item Turn to another user in $\cal{M}$.
\end{enumerate}
Iterations will not stop \emph{\textbf{until}} no user can form a swap-blocking pair with any other users in a new round.

\textbf{Step 3: \emph{End of algorithm.}}\\
\hline

\end{tabular}
\end{table}

Note that USMA-1 is not guaranteed to converge to a global optimal\footnote{An optimal matching refers to a matching reaching the global maximum utility of the network.} 2ES matching, and we will explain that in Section \uppercase\expandafter{\romannumeral 4}.C. in detail. We then propose USMA-2 to search the global optimal matching based on a simulated annealing method~\cite{L-1987}. In USMA-2, we start with a random initial matching. In the swap matching phase, we do not care whether the swap matching is approved any more, instead, a swap matching $\Psi _{jq}^{ip}$ is executed with a probability $P_T$ which depends on the total sum-rate as shown below:
\begin{equation} \label{PT}
{P_T} = \frac{1}{{1 + {e^{ - T\left[ {{U_{total}}\left( {\Psi _{jq}^{ip}} \right) - {U_{total}}\left( \Psi  \right)} \right]}}}},
\end{equation}
where $T$ is a probability parameter. The algorithm keeps tracking the optimal matching found so far, even if the utility of current matching is not a local maximum. The details of USMA-2 is presented in Table \ref{alg-SM-2}, and specific analysis can be found in Section \uppercase\expandafter{\romannumeral 4}.C.

\begin{table}[!t]
\renewcommand{\arraystretch}{1.0}
\caption{User-Subchannel Matching Algorithm (USMA-2)}
\label{alg-SM-2}
\centering
\begin{tabular}{p{150mm}}
\hline

\textbf{Step 1: \emph{Initialization Phase}}

\begin{enumerate}
    \item Record current matching as $\Psi$.
    \item Users and sub-channels are randomly matched with each other subject to $\left| {\Psi \left( {S{C_k}} \right)} \right| \le {d_f}$ and $\left| {\Psi \left( {{M_j}} \right)} \right| \le {d_v}$.
    \item Set $U_{max} = U_{total}\left( \Psi \right)$.
\end{enumerate}

{\bf Step 2: \emph{Swap matching phase}.}

\textbf{while} $\ell \le {\ell_{max}}$,
\begin{enumerate}
\item Randomly select a pair of users $\left( {{M_i},{M_j}} \right)$ and sub-channels $\left( {{{SC}_p},{{SC}_q}} \right)$ such that $S{C_p} \in \Psi \left( {{M_i}} \right)$, $S{C_q} \in \Psi \left( {{M_j}} \right)$, and $S{C_p} \notin \Psi \left( {{M_j}} \right)$, $S{C_q} \notin \Psi \left( {{M_i}} \right)$.
\item Calculate $P_T$ according to equation $\left( \ref{PT} \right)$.
\item Execute swap-matching ${\Psi _{jq}^{ip}}$, and set $\Psi = {\Psi _{jq}^{ip}}$ with probability $P_T$.
\item If $U_{total}\left( {\Psi _{jq}^{ip}} \right) > U_{max}$, then set $U_{max} = U_{total}\left( {\Psi _{jq}^{ip}} \right)$.
\item $\ell = \ell + 1$.
\end{enumerate}
\textbf{end while}.

\textbf{Step 3: \emph{End of algorithm.}}\\
\hline

\end{tabular}
\end{table}

With the above two sub-channel allocation algorithms, we can then present the overall resource allocation algorithm for the problem in $\left( \ref{system_optimization} \right)$. In the initialization phase, the BS allocates the transmitted power equally to each user over each sub-channel, and the weight factor $w_j$ for each user $M_j$ is set as inversely proportional to the average rate of user $M_j$ in previous time slots. In the resource allocation phase, sub-channel assignment and power allocation are iteratively performed so as to obtain a joint solution.

\begin{table}[!t]
\renewcommand{\arraystretch}{1.0}
\caption{Joint Subchannel and Power Allocation Algorithm (JSPA)}
\label{algt}
\centering
\begin{tabular}{p{150mm}}
\hline

\textbf{Step 1: \emph{Initialization Phase}}

\begin{enumerate}
    \item The BS obtains CSI of all the users.
    \item The BS allocates the transmitted power equally to each user over each sub-channel.
    \item Set ${w_j} = a/{{\bar R}_j}$ for any $M_j \in \cal{M}$, in which $a$ is the inverse scaling factor, and ${\bar R}_j$ is average data rates of user $M_j$ in previous time slots.
    \item Set $t = 0$.
\end{enumerate}

{\bf Step 2: \emph{Joint Sub-channel and Power Allocation}.}

\textbf{repeat}
\begin{enumerate}
\item Update the sub-channel allocation matrix $\emph{\textbf{B}}$ by solving the matching problem in Definition 1 using USMA-1 or USMA-2.
\item Update $\emph{\textbf{p}}$ by solving GP formulated in $\left( \ref{power_GP} \right)$ using the interior point methods.
\item Set $t = t + 1$.
\end{enumerate}
\textbf{until} convergence.

\textbf{Step 3: \emph{End of algorithm.}}\\
\hline

\end{tabular}
\end{table}

\subsection{Stability, Convergence, Complexity and Optimality}
Given the proposed USMA-1 and USMA-2 above, we then give remarks on the stability, convergence, complexity, and optimality.

\subsubsection{Stability and convergence}
We now prove the stability and convergence of USMA-1 and JSPA (with $t_{max} = + \infty$), while the convergence of USMA-2 is usually not considered as it is usually constrained by the maximum iteration number $\ell_{max}$.

\textbf{Lemma 1:} If USMA-1 converges to a matching $\Psi^*$, then $\Psi^*$ is a \emph{\textbf{2ES}} matching.

\begin{proof}
According to Table \ref{alg-SM-1}, when the proposed USMA-1 converges to a terminal matching $\Psi^*$, any user $M_j \in {\cal{M}}$ cannot find another user $M_i \in {\cal{M}}$ to form a swap-blocking pair along with their matches. Thus, the matches of user $M_j$ must be the best choice for it in current matching. There is no user that can improve its utility by a unilateral change of its matches. Hence, the terminal matching $\Psi^*$ is \emph{\textbf{2ES}}.
\end{proof}

\textbf{Theorem 1:} The proposed USMA-1 converges to a \textbf{\emph{2ES}} matching ${\Phi ^*}$ after a limited number of swap operations.

\begin{proof}
Convergence of USMA-1 depends on Step 2 in Table \ref{alg-SM-1}. After a number of swap operations, the structure of matching changes as follows:
\begin{equation}
{\Psi _0} \to {\Psi _1} \to {\Psi _2} \to  \cdots.
\end{equation}
After swap operation ${\ell}$, the matching changes from ${\Psi _{\ell  - 1}}$ to ${\Psi _{\ell}}$. Without loss of generality, we assume that the pair of users resulting in this swap-matching is $\left(M_i, M_j \right)$ with ${\Psi _\ell } = {\Psi _{\ell  - 1}}_{jq}^{ip}$. According to Definition 3, after each swap operation, the utility of ${SC}_p$ and ${SC}_q$ satisfies ${R_{S{C_p}}}\left( {{\Psi _\ell }} \right) \ge {R_{S{C_p}}}\left( {{\Psi _{\ell  - 1}}} \right)$ and ${R_{S{C_q}}}\left( {{\Psi _\ell }} \right) \ge {R_{S{C_q}}}\left( {{\Psi _{\ell  - 1}}} \right)$, in which at least one of the equalities does not stand. The utilities of other sub-channels keep the same. Therefore, the total sum-rate over all the sub-channel increase after each swap-matching operation ${\ell}$:
\begin{equation} \label{iteration_sumrate}
\Delta _{\ell  - 1}^\ell : = {U_{total}}\left( {{\Psi _\ell }} \right) - {U_{total}}\left( {{\Psi _{\ell  - 1}}} \right) = {\sum _{k \in {{\cal K}}}}{R_{S{C_k}}}\left( {{\Psi _\ell }} \right) - {\sum _{k \in {{\cal K}}}}{R_{S{C_k}}}\left( {{\Psi _{\ell  - 1}}} \right) > 0.
\end{equation}
Note that the number of potential swap-blocking pairs is finite since the number of matched users is limited, and the total sum-rate has an upper bound due to limited spectrum resources. Therefore, there exists a swap operation ${\ell}^*$ after which there exists no approved swap operation and the total sum-rate stops increasing. USMA-1 then converges to a final matching ${\Phi ^*}$, which is a \emph{\textbf{2ES}} matching according to Lemma 1.
\end{proof}

\textbf{Theorem 2:} The proposed JSPA for resource allocation is guaranteed to converge.

\begin{proof}
The proof is analogous to that of Theorem 1. After a number of iterations, the total sum rates change as follows:
\begin{equation}
{U_{total} ^0} \to {U_{total} ^1} \to {U_{total} ^2} \to  \cdots,
\end{equation}
in which after iteration $t$, the total sum rates change from ${U_{total} ^{t-1}}$ to $U_{total} ^{t}$. Each iteration $t$ of JSPA consists of two phases: USMA and power allocation. In Theorem 1, we have proved that the total sum-rate will increase after USMA-1 is performed. Even if there is no approved swap matching, the total sum-rates remain the same. From Table \ref{alg-SM-2}, it is guaranteed that at least the total sum-rate will not decrease after USMA-2 is performed. We assume that the matching at the beginning of iteration $t$ is $\Psi_t$ and the matching obtained at the end of iteration $t$ is $\Psi'_t$, then the following stands:
\begin{equation}
{U_{total}}\left( {{{\Psi '}_t}} \right) \ge {U_{total}}\left( {{\Psi _t}} \right).
\end{equation}

Based on the convex optimization problem $\left( \ref{power_GP} \right)$ formulated for power allocation, we can see that the total sum-rate will increase after Step-2-3 and Step 2-4 in Table \ref{algt} are executed, unless the initial power allocation scheme is exactly the solution for $\left( \ref{power_GP} \right)$. Therefore, in each iteration of JSPA, the total sum-rate grows after both the sub-channel and power allocation, i.e.,
\begin{equation}
{U_{total}^t} > {U_{total}^{t-1}}.
\end{equation}
Since there exists an upper bound for the total sum-rate, it will stop increasing after a limited number of iterations in JSPA, and then the algorithm converges.
\end{proof}

\subsubsection{Complexity}
Given the convergence of the proposed USMA-1, we can then discuss the computational complexity of USMA-1. For the initialization phase, the complexity mainly lies in the process of sorting the users' weights, which is ${\rm O}\left( {{M^2}} \right)$ in average. Note that in the swap-matching phase, a number of iterations are operated to reach the final matching. In every iteration, the BS searches for swap-blocking pairs and the users execute all the approved swap operations over two corresponding sub-channels. So the complexity of the swap-matching phase lies in the number of both iterations and attempts of swap matchings in each iteration.

\textbf{Proposition 4:} In each iteration of USMA-1, at most $\frac{1}{2}M{d_f}{d_v}\left( {K - {d_v}} \right)$ swap matchings need to be considered when $M{d_v} = K{d_f}$. Given the number of total iterations $I$, the computational complexity of USMA-1 can be approximated as $O\left( {IM{d_f}{d_v}K} \right)$.

\begin{proof}
When $M{d_v} = K{d_f}$, each player remains fully matched before and after every swap matching, and thus, any swap matching ${\Psi _{jq}^{ip}}$ consists of two actual users and two sub-channels. For user $M_i$, there exist ${d_v}\left(K - d_v \right)$ possible combinations of ${SC}_p$ and ${SC}_q$ in ${\Psi _{jq}^{ip}}$ since there are $K$ sub-channels and each user can occupy at most $d_v$ ones. For the chosen ${SC}_q$, at most $d_f$ possible $M_j$ need to be considered. Therefore, a swap matching ${\Psi _{jq}^{ip}}$ with $M_i$ fixed has ${d_f}{d_v}\left( {K - {d_v}} \right)$ possible combinations. Since there are $M$ users, at most $\frac{1}{2}M{d_f}{d_v}\left( {K - {d_v}} \right)$ swap matchings need to be considered in each iteration of USMA-1. In practice, one iteration requires a significantly low number of swap operations, since the values of $d_v$ and $d_f$ are usually rather small. Therefore, given the number of total iterations $I$, the computational complexity of USMA-1 can be presented by $O\left( {IM{d_f}{d_v}K} \right)$.
\end{proof}

Note that the total number of iterations in USMA-1 and JSPA cannot be given in closed form since we don't know for sure at which iteration the users form a 2ES matching or the total sum-rate stops increasing, which is common in the design of most heuristic algorithms. To evaluate the convergence, we will show the distribution of the total number of swap matchings required for USMA-1 in Fig.~\ref{alphatwo:a} and that of the number of iterations in the JSPA, i.e., $t$, in Fig.~\ref{alphatwo:b}. Corresponding analysis will be given in Section~\uppercase\expandafter{\romannumeral5}.

\subsubsection{Optimality}
We show below whether USMA-1 and USMA-2 can achieve an optimal matching, and that the global optimal solution of resource allocation problem in $\left( \ref{system_optimization} \right)$ can be obtained by utilizing the proposed JSPA with USMA-2 applied.

\textbf{Theorem 3:} All local maxima of $U_{total}$ corresponds to a 2ES matching.

\begin{proof}
Suppose the total utility of matching $\Psi$ is a local maximum of $U_{total}$. If $\Psi$ is not a 2ES matching, then any approved swap matching strictly increases $U_{total}$ according to Theorem 1. However, this is in contradiction to the assumption that $\Psi$ is a local maximum. Therefore, $\Psi$ must be 2ES.
\end{proof}

However, not all 2ES matchings obtained from USMA-1 are local maxima of $U_{total}$. For example, there exists possibility that a user $M_i$ does not approve a swap operation ${\Psi _{jq}^{ip}}$ since its utility will decrease, but another user $M_j$ will benefit a lot from this swap operation, and the utility of ${SC}_p$ and ${SC}_q$ will increase. If the swap operation is \emph{forced}, then the total utility will increase at the expense of a weaker stability, as expressed in the following remark.

\textbf{Remark 2:} In USMA-1, a forced swap matching will further increase the total utility compared to an approved swap matching, resulting in a one-sided exchange stable matching.

\textbf{Proposition 5:} With sufficiently large $\ell_{max}$, USMA-2 reaches a global optimum of the total utility, which is also a 2ES matching.
\begin{proof}
USMA-2 is proposed based on the simulated annealing algorithm, which has been proved to reach a global optimum with a sufficiently large number of iterations in~\cite{O-2001}\cite{L-1987}. For a global optimum, there is no approved swap matching that can further improve the total utility, i.e., there is no swap matching to improve a player's utility without hurting others'. Therefore, it is natural that this is also a 2ES matching.
\end{proof}

\textbf{Proposition 6:} For a sufficiently large $\ell_{max}$, JSPA (USMA-2 applied) reaches a local optimal solution for the resource allocation problem in $\left( \ref{system_optimization} \right)$.

Since the computational complexity of USMA-2 is usually extremely high (approximately exponential time), we set a fixed value of $\ell_{max}$ in USMA-2.

%%%%%%%%%%%%%%%%%%%%%%%%%%%%%
\section{Simulation Results}%
%%%%%%%%%%%%%%%%%%%%%%%%%%%%%

%\begin{figure}[!t]
%\centering
%\includegraphics[width=5in]{iteration_CDF.eps}
%\caption{CDF of the total number of swap operations in USMA-1 v.s. number of swap operations.} \label{iteration_CDF}
%\end{figure}
%
%\begin{figure}[!t]
%\centering
%\includegraphics[width=5in]{JSPA_iteration_CDF.eps}
%\caption{CDF of the total number of iterations $t$ v.s. number of iterations $t$ in JSPA (USMA-1 applied).} \label{JSPA_iteration_CDF}
%\end{figure}

\begin{figure}
\centering
\subfigure[C.D.F. of the number of swap operations in USMA-1]{
\label{alphatwo:a} %% label for first subfigure
\includegraphics[width=3.15in]{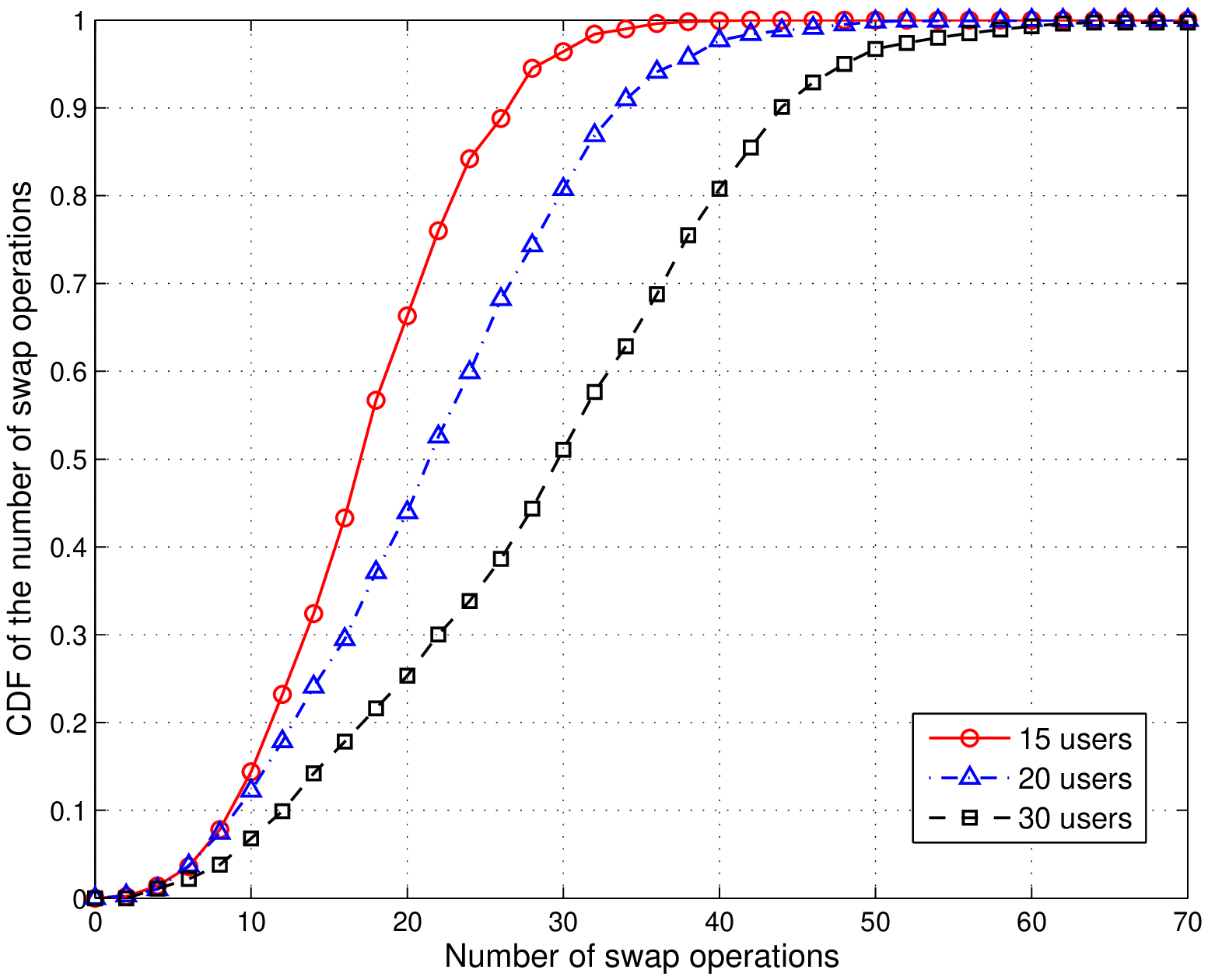}}
\hspace{-0.2in}
\subfigure[C.D.F. of the total number of iterations $t$ in JSPA (USMA-1 applied)]{
\label{alphatwo:b} %% label for second subfigure
\includegraphics[width=3.3in]{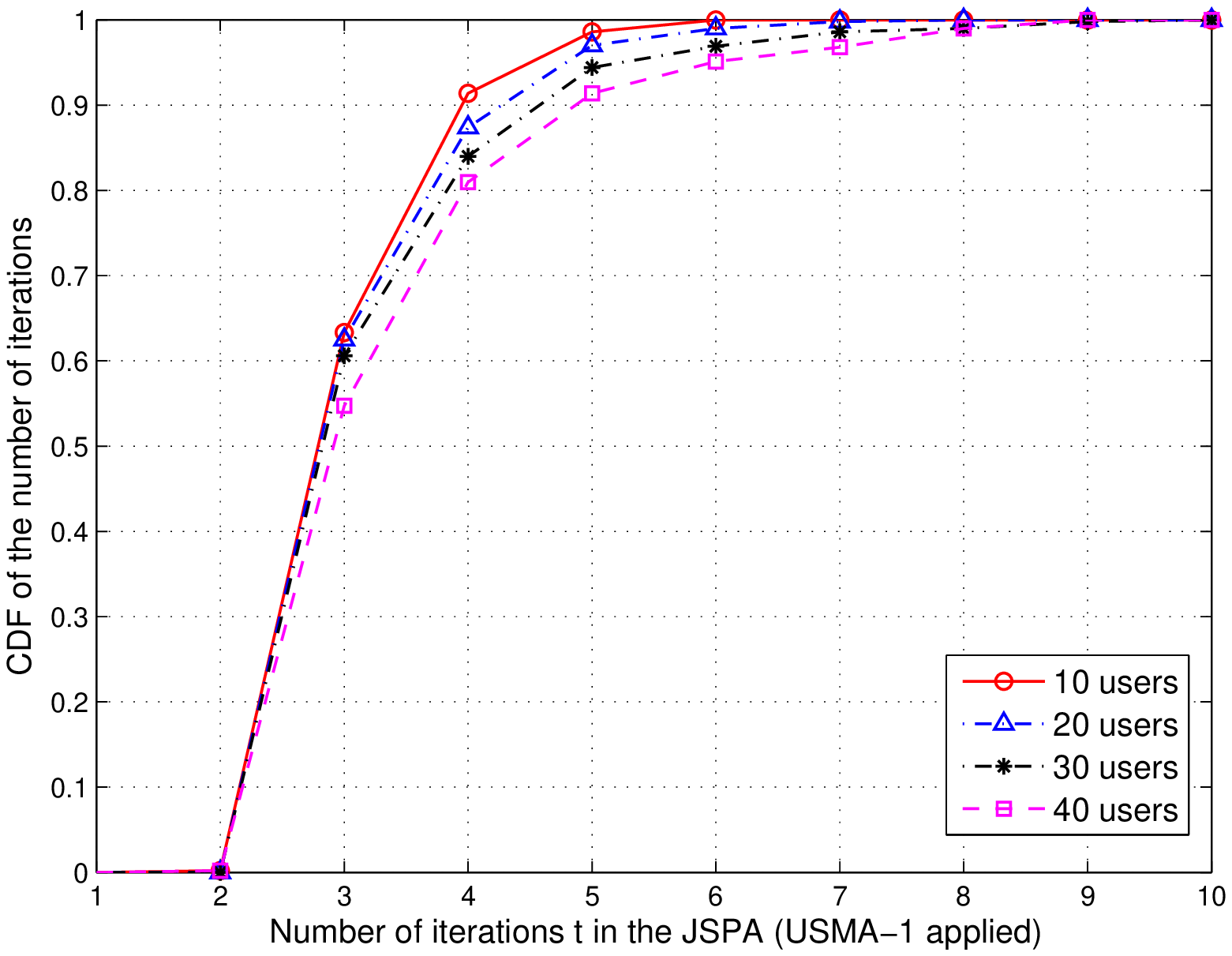}}
\caption{Distribution of the total number of swap operations in USMA-1 and that of the total number of iterations $t$ in JSPA (USMA-1 applied)} \label{alphatwo} %% label for entire figure
\end{figure}

\begin{figure}[!t]
\centering
\includegraphics[width=5in]{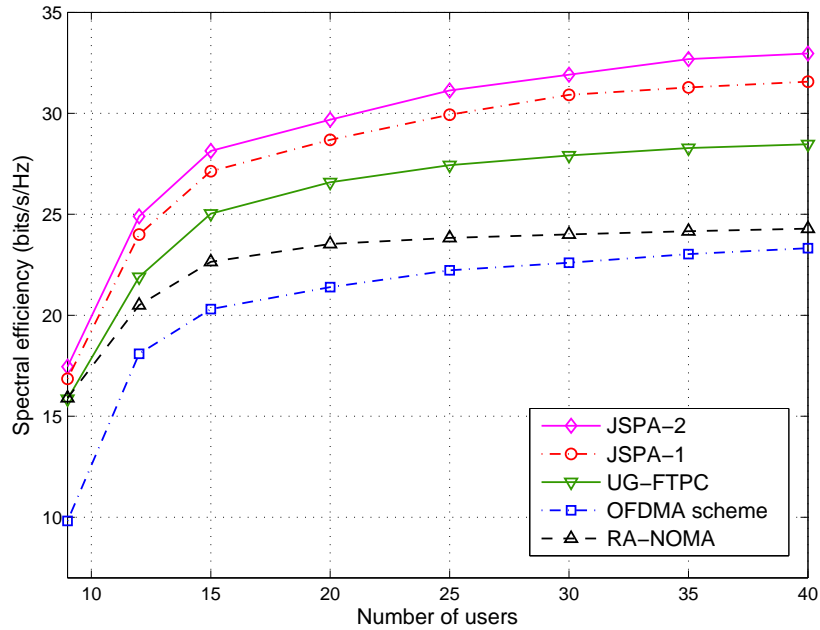}
\caption{Spectral efficiency vs. number of the users.} \label{rate_user_num}
\end{figure}

\begin{figure}[!t]
\centering
\includegraphics[width=5in]{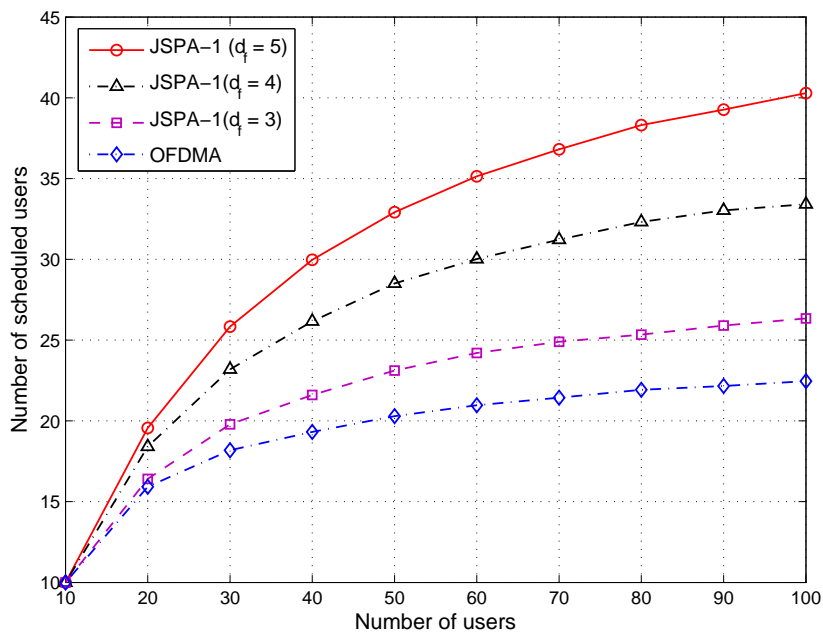}
\caption{Number of scheduled users vs. number of the users.} \label{user_user_num}
\end{figure}

\begin{figure}[!t]
\centering
\includegraphics[width=5in]{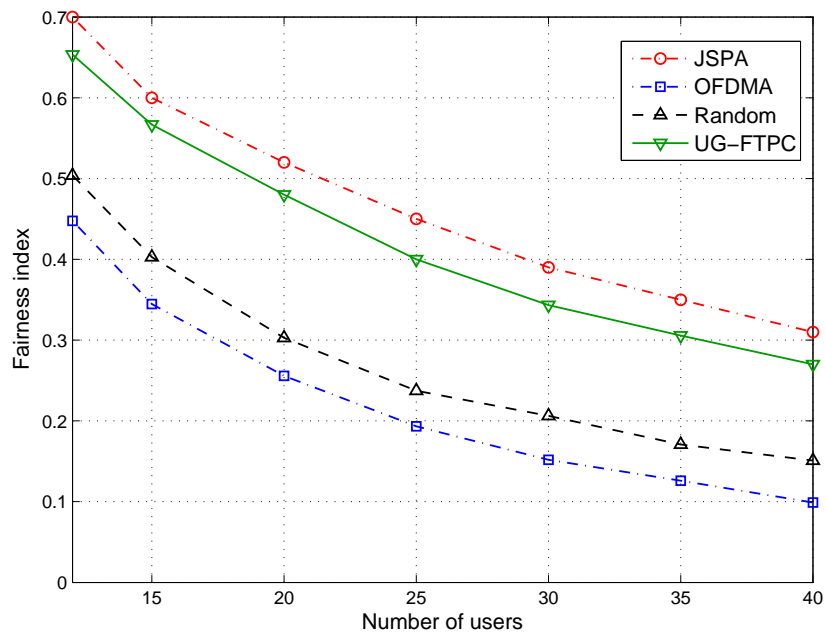}
\caption{User fairness index vs. number of the users.} \label{user_fairness}
\end{figure}

\begin{figure}[!t]
\centering
\includegraphics[width=5in]{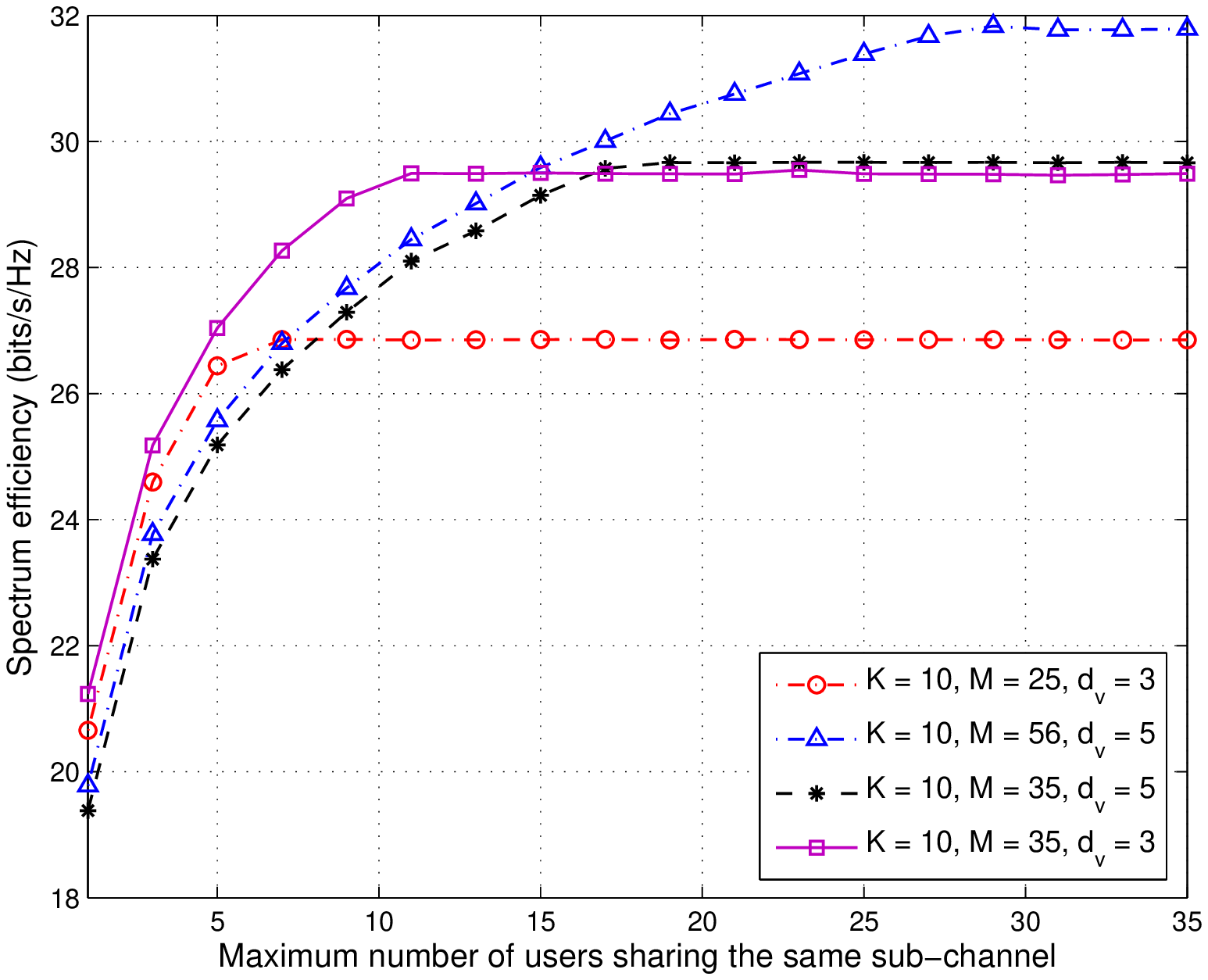}
\caption{Spectrum efficiency v.s. maximum number of users sharing the same sub-channel.} \label{df_rate}
\end{figure}

In this section, we evaluate the performance of the proposed JPSA with both USMA-1 and USMA-2 applied, and compare its performance with the OFDMA scheme and a random allocation scheme (RA-NOMA). In the OFDMA scheme, we assume that each sub-channel can only be assigned to one user, and joint sub-channel and power allocation is performed by utilizing the utility-based dynamic algorithm in~\cite{GL-2003}. In the RA-NOMA scheme, the set of sub-channels is randomly allocated to the users satisfying constraints $\left( \ref{system_5} \right)$ and $\left( \ref{system_7} \right)$. We set ${\ell _{max}} = 2 \times {10^6}, T = 0.5$ in USMA-2. For convenience, we refer to the JSPA with USMA-2 as JSPA-2, and the JSPA with USMA-1 as JSPA-1. To better evaluate the performance of our proposed algorithms, a previous resource allocation algorithm in~\cite{AAYYAT-2013} based on user grouping and fractional transmit power control (UG-FTPC) is adopted. In the UG-FTPC method, the users are separated as $d_v$ groups according to their channel gains, and each user can only share subcarriers with the users who are not in the same group with it. For the power allocation method FTPC, more power is allocated to the users with inferior channel condition for the fairness consideration~\cite{AAYYAT-2013}. For the simulations, we set the BS's peak power, $P_s$ to 46dBm, noise power spectral density to -174 dBm/Hz, carrier center frequency to 2GHz, system bandwidth to 4.5MHz based on existing LTE/LTE-Advanced specifications~\cite{3GPP25,3GPP36}. For the OFDMA scheme, the total bandwidth is divided into 25 sub-channels, while for the NOMA scheme, we set the number of sub-channels as 10 considering the decoding complexity and signaling cost for the receivers at the BS. We assume that the pass loss is obtained by a modified Hata urban propagation model~\cite{3GPP25}, and that all users are uniformly distributed in a square area with the size of length $350$m. Simulation results are obtained as shown below.

Denote the random variable $\tilde Y$ as the total number of swap operations required for USMA-1 to converge. Fig.~\ref{alphatwo:a} shows the cumulative distribution function (C.D.F.) of $\tilde Y$, $\Pr \left( {\tilde Y \le \tilde y} \right)$, versus $\tilde y$ for different number of users, with the number of sub-channels $K = 10$, maximum number of sub-channels that a user has access to simultaneously $d_v = 4$, and maximum number of users sharing the same sub-channel $d_f = 3$. We observe that the speed of convergence becomes faster as the number of the users decreases. Besides, Fig.~\ref{alphatwo:a} further reflects that the computational complexity is rather low in the proposed USMA-1. For example, when the number of the users is 30, on average a maximum of 70 iterations are needed for USMA-1 to converge. Similarly, Fig.~\ref{alphatwo:b} shows the C.D.F. of the total number of iterations required for the JSPA-1. As seen from Fig.~\ref{alphatwo:b}, the proposed JSPA-1 converges within 3-10 iterations, depending on the operating scenario.

Fig.~\ref{rate_user_num} illustrates the spectrum efficiency vs. the number of users $M$ with $d_f = 3, d_v = 5$ in the NOMA scheme. We evaluate the spectrum efficiency by obtaining the average total sum-rate within 30 slots. We find out that the spectrum efficiency increases with the number of users, and the rate growth becomes slower as $M$ increases. When the number of users is much larger than the number of sub-channels, the total sum-rate continues to increase due to the multiuser diversity gain but grows at a slower speed. This makes sense since the influence of multiuser diversity is more significant when the number of users is small. From Fig.~\ref{rate_user_num}, we can see that the proposed JSPA performs much better than the OFDMA scheme and two NOMA schemes, i.e., the RA-NOMA and the UG-FTPC. This is because our proposed JSPA provides more freedom in the subcarrier allocation than the predefined user grouping strategy in the UG-FTPC. Our proposed JSPA-2 provides higher spectral efficiency than the JSPA-1 at the expense of high complexity. Fig.~\ref{rate_user_num} further implies that the BS does not take full advantage of the spectrum resources in the OFDMA scheme since one sub-channel can only be assigned to one user.

Fig.~\ref{user_user_num} shows the average number of scheduled users v.s. the number of users with $d_v = 5$ in the NOMA scheme within 30 slots. When the number of users is smaller than or equal to the number of sub-channels, all the users have access to the spectrum resources in both OFDMA and NOMA schemes. As the number of users exceeds the number of sub-channels, theoretically only up to 25 users can access the spectrum resources simultaneously in the OFDMA scheme. In practice, the number of scheduled users is smaller than 25 since one user may be accessed to more than one sub-channels. Thus in the OFDMA scheme, user connectivity drops badly when there are large number of users in one cell. In the NOMA scheme, as the number of users grows, the number of scheduled users tends to be a fixed value which is smaller than $K{d_f}$ but is still much larger than that of the OFDMA scheme. Note that the number of scheduled users is higher when $d_f$ becomes larger, since more users have the opportunity to be served by the BS.

Fig.~\ref{user_fairness} shows the user fairness index v.s. the number of users with $d_f = 3$ and $d_v = 5$ in the NOMA scheme. To evaluate the user fairness, we record the average sum-rate of each user $M_j$ within 30 slots as ${\bar R_j}$. Following the setup of~\cite{YYATAK-2013}, we introduce Jain's fairness index~\cite{RDW-1984} which can be calculated as ${\left( {\sum\nolimits_{j \in \mathcal{M}} {{{\bar R}_j}} } \right)^2}/\left( {K\sum\nolimits_{j \in \mathcal{M}} {\bar R_j^2} } \right)$. The value range of Jain's fairness is between 0 and 1 with the maximum achieved by equal users' rates. From Fig.~\ref{user_fairness}, we observe that the fairness index decreases as the number of users increases since the competition between users is tenser. In addition, the proposed JSPA achieves higher user fairness than other NOMA schemes and the OFDMA scheme, implying that the NOMA scheme has more potential than traditional OFDMA scheme in achieving massive connectivity.

Fig.~\ref{df_rate} depicts the total sum-rate v.s. maximum number of users sharing the same sub-channel, $d_f$, in the NOMA scheme. With different settings of $M$ and $d_v$, the total sum-rate grows to a stable value as $d_f$ increases, because each user has been matched to at most $d_v$ sub-channels after $d_f$ reaches $N{d_v}/K$. Note that when $d_f = 1$, this is an OMA scheme with lower spectral efficiency than the NOMA schemes in which $d_f > 1$. Considering the computational complexity of SIC which grows with $d_f$, we make an initial observation that the value of $d_f$ should be smaller than $d_f^*$ so as to reach a balance between the spectrum efficiency and decoding complexity, in which $d_f^*$ is the value of inflection point in each curve.

%%%%%%%%%%%%%%%%%%%%%%
\section{Conclusion}%
%%%%%%%%%%%%%%%%%%%%%%
In this paper, we studied the resource allocation problem in a downlink NOMA wireless network by jointly optimizing the sub-channel assignment and the power allocation, while achieving a balance between user fairness and the maximization of the total sum-rate. By formulating the sub-channel allocation problem as a many-to-many two-sided matching problem with externalities, we proposed a low-complexity user-subchannel swap matching algorithm in which the users and sub-channels can be matched and form a two-sided exchange stable matching. Properties of the proposed algorithm have been discussed including the global and local optimality. A tradeoff can be reached between the total sum-rate and the decoding complexity by setting the value of $d_f$, which represents the maximum number of users sharing the same sub-channel. The NOMA scheme outperforms the traditional OFDMA scheme in terms of both the total sum-rate and the user fairness.

\begin{appendices}

\section{}
\emph{Proof of Proposition 1:} The proof can be separated into two cases in which $d_f = 1$ and $d_f > 1$, respectively.
\begin{enumerate}
\item When $d_f = 1$, $\left( \ref{system_optimization} \right)$ becomes an joint power and sub-channel allocation problem in the traditional OFDMA system, which has been proved to be NP-hard in~\cite{YY-2014}.
\item When $d_f > 1$, we prove that the problem is NP-hard even when power allocation is not considered. We construct an instance of $\left( \ref{system_optimization} \right)$ and establish the equivalence between this instance and a three-sided matching problem. We consider an instance in which $d_f = 2$, and $d_v = 1$. Suppose the BS allocates the power to each active user over every sub-channel equally, and the users are separate into two disjoint sets ${\cal{M}}_1$ and ${\cal{M}}_2$ such that ${\mathcal{M}_1} \cup {\mathcal{M}_2} = \mathcal{M}$ and ${M_1} \cap {M_2} = \emptyset$. Each ${SC}_k$ is allocated to one user from ${\cal{M}}_1$ and another user from ${\cal{M}}_1$. Then the instance becomes a sub-channel allocation problem. We define the decision problem of it and reduce the decision problem to a 3-dimensional matching problem (3-DM problem). Since the 3-DM problem has been proven to be NP-comlete \cite{5K-1972,55K-1991}, the decision problem of this instance is also NP-complete. Thus, the instance of $\left( \ref{system_optimization} \right)$ with equal power allocation is an NP-hard problem~\cite{33S-2012}.
    \begin{enumerate}
        \item Let's first obtain the decision problem of the instance with equal power allocation. Let ${\cal{M}}_1$, ${\cal{M}}_2$, and ${\mathcal{K}}$ be three disjoint sets of users, and sub-channels, respectively. We have $\left| {{{\cal{M}}_1}} \right| = M/2$, $\left| {{{\cal{M}}_2}} \right| = M/2$, and $\left| {\mathcal{K}} \right| = K$. Let $\cal{Q}$ be a collection of ordered triples ${\cal{Q}} \subseteq {\mathcal{K}} \times {{\cal{M}}_1} \times {{\cal{M}}_2}$, where ${\textbf{\emph{Q}}_i} = \left( {S{C_k},M_i,M_j} \right) \in {\cal{Q}}$. According to $\left( \ref{rate_subchannel} \right)$, the sum-rate of any triple ${\textbf{\emph{Q}}_i}$ can be set as ${R_{{\textbf{\emph{Q}}_i}}}$. To be convenient, set $L = \min \left\{ {M/2,K} \right\}$. Now we need to determine whether there exists a set ${\cal{Q}}' \subseteq {\cal{Q}}$ so that $\left| {{\cal{Q}}'} \right| = L$, $\sum\limits_{i = 1}^L {{S_{{\textbf{\emph{Q}}_i}^\prime }}}  \ge \lambda $, where any ${\textbf{\emph{Q}}_i}^\prime  \in \mathcal{Q'}$ and ${\textbf{\emph{Q}}_j}^\prime  \in \mathcal{Q'}$ do not contain the same components.
        \item Next let's present a traditional 3-DM decision problem. Let ${\cal{M}}_1$, ${\cal{M}}_2$, and ${\mathcal{K}}$ be three disjoint sets of users and sub-channels, respectively. Let $\cal{Q}$ be a collection of ordered triples ${\cal{Q}} \subseteq {\mathcal{K}} \times {{\cal{M}}_1} \times {{\cal{M}}_2}$. Then ${\mathcal{Q'}} \subseteq {\cal{Q}}$ is a 3-DM if the followings hold: 1)$\left| {{\cal{Q}}'} \right| = L$; 2)for any two distinct triples $\left( {S{C_i},M_i,M_j} \right) \in {{\cal{Q}}'}$ and $\left( {S{C_p}, M_p, M_q} \right) \in {{\cal{Q}}'}$, we have $i \ne j \ne p \ne q$. It has been shown that a 3-DM decision problem is an NP-complete problem even in the special case that $\left| M/2 \right| = \left| K \right|$~\cite{6ACGKMP-2003}.
        \item We then show that the problems in a) and b) are equivalent. For the decision problem formulated in a), if let $\lambda$ go to an infinite negative, the decision problem of the above instance can be reduced to a 3-DM decision problem. Therefore, the decision problem in a) is NP-complete, and the corresponding instance is NP-hard.
    \end{enumerate}
    A special case of $\left( \ref{system_optimization} \right)$ is NP-hard, and Proposition 1 stands. $\hfill\blacksquare$
\end{enumerate}

\vspace{0.4cm}
\section{}
\emph{Proof of Proposition 2:} If we remove constraint $\left( \ref{system_7} \right)$ and rewrite $\left( \ref{system_1} \right)$, then the problem in $\left( \ref{system_optimization} \right)$ is equivalent to a single-subchannel resource allocation problem, since the power constraints are separate over each sub-channel. The optimal solution for the BS is to allocate each sub-channel ${SC}_k$ to only one user $M_{j^*}$ with transmitted power $P_k$ satisfying ${j^*} = \arg \mathop {\max }\limits_{j \in \cal{M}} \left( {{{\left| {{h_{k,j}}} \right|}^2}/{n_{k,j}}} \right)$. We give a simple example below to explain that if any other user $M_i \in \cal{M}$ is accessed to ${SC}_k$, the data rates of ${SC}_k$ will drop.

\textbf{Example:} since the channel gain of $M_{j^*}$ is the largest among all the users, we have${\left| {{h_{k,{j^*}}}} \right|^2}/{n_{j^*}} > {\left| {{h_{k,i}}} \right|^2}/{n_{k,i}}$. If they share sub-channel ${SC}_k$, the sum-rate produced over this sub-channel is presented as
\begin{equation} \label{sum-rate_ij}
{R_{i+{j^*}}} = {\log _2}\left( {1 + \frac{{\beta {P_k}{{\left| {{h_{k,{j^*}}}} \right|}^2}}}{{{n_{j^*}}}}} \right) + {\log _2}\left( {1 + \frac{{\left( {1 - \beta } \right){P_k}{{\left| {{h_{k,i}}} \right|}^2}}}{{\beta {P_k}{{\left| {{h_{k,i}}} \right|}^2} + {n_{k,i}}}}} \right),
\end{equation}
where $\beta$ is the proportional factor of power allocation. If only user ${j^*}$ occupies this channel, then the sum-rate over this sub-channel is
\begin{equation} \label{sum-rate_j}
{R_{j^*}} = {\log _2}\left( {1 + \frac{{{P_K}{{\left| {{h_{k,{j^*}}}} \right|}^2}}}{{{n_{k,j^*}}}}} \right).
\end{equation}
We can easily derive that ${R_{j^*}} > {R_{i + {j^*}}}$.

The above two-user case can be extended to a multi-user single-subchannel one, and thus, the optimal solution for the relaxed version of problem $\left( \ref{system_optimization} \right)$ is actually a OMA resource allocation scheme. However, for the general version $\left( \ref{system_optimization} \right)$, this proposition does not stand any more. $\hfill\blacksquare$

\section{}
\emph{Proof of Proposition 3:} Given problem $\left( \ref{power_GP} \right)$, we follow the method in~\cite{S-2008} to convert it into GP\footnote{The optimal decoding order, i.e., the SIC decoding in this case, determines that this problem can be successfully converted into GP~\cite{S-2008}.}. The achievable rate region over sub-channel ${SC}_k$ can be represented as
\begin{equation} \label{power_conv}
\textbf{\emph{R}}\left( {m\left( k \right),\left\{ {{p_{k,j}}} \right\}} \right) = \left\{ {{R_{k,{\pi _k}\left( j \right)}}:{R_{k,{\pi _k}\left( j \right)}} \le \log \left( {1 + \frac{{{p_{k,{\pi _k}\left( j \right)}}}}{{{m_{k,{\pi _k}\left( j \right)}} + \sum\nolimits_{i < j} {{p_{k,{\pi _k}\left( i \right)}}} }}} \right),j = 1, \cdots ,{d_f}} \right\}
\end{equation}
When each rate vector for sub-channel ${SC}_k$, i.e., ${\left\{ {{R_{k,{\pi _k}\left( j \right)}}} \right\}_{1 \times {d_f}}}$, reaches the boundary of the capacity region, the following equation stands for obtaining the power sets ${\left\{ {{p_{k,j}}} \right\}}$ with rate vector as variables:
\begin{equation} \label{power_change}
\sum\limits_{i = 1}^t {{p_{k,{\pi _k}\left( i \right)}}}  = \sum\limits_{i = 1}^t {\left( {{m_{k,{\pi _k}\left( i \right)}} - {m_{k,{\pi _k}\left( {i - 1} \right)}}} \right)}  \times {e^{\sum\nolimits_{j = 1}^t {{R_{k,{\pi _{_k}}\left( j \right)}}} \ln 2}} - {m_{k,{\pi _k}\left( t \right)}},t = 1,2, \cdots ,{d_f}
\end{equation}
Substituting equation $\left( \ref{power_change} \right)$ into $\left( \ref{power_conv} \right)$, we see that equation $\left( \ref{power_conv} \right)$ is then equivalent to
\begin{equation}
\begin{aligned}
R\left( {m\left( k \right),\left\{ {{p_{k,j}}} \right\}} \right) = & \left\{ {{{R_{k,{\pi _k}\left( j \right)}}:\sum\limits_{i = 1}^{{d_f}} {\left( {{m_{k,{\pi _k}\left( i \right)}} - {m_{k,{\pi _k}\left( {i - 1} \right)}}} \right)}  \times {e^{\sum\nolimits_{j = i}^{{d_f}} {{R_{k,{\pi _{_k}}\left( j \right)}}} \ln 2}} \le }}\right.\\
& \left. { {\sum _{j \in {{\cal S}_k}}}{p_{k,j}} + {m_{k,{\pi _k}\left( {{d_f}} \right)}},j = 1, \cdots ,{d_f} } \right\}
\end{aligned}
\end{equation}
Therefore, problem $\left( \ref{power_allocation} \right)$ can be converted into $\left( \ref{power_GP} \right)$ with rate vector as optimization variables. $\hfill\blacksquare$

\section{}
\emph{Proof of Remark 1:} Since each user can be matched with more than one sub-channel, and each sub-channel can be matched with a subset of users, this is a many-to-many matching game. Due to the interference item $\left( \ref{interference_user} \right)$ in equation $\left( \ref{throughput_single}\right)$, any user $M_j$'s sum rate over its occupied sub-channel, say, ${SC}_k$, is related to the set of other users sharing this sub-channel, i.e., $\mathcal{S}_k$. Thus, each user cares not only which sub-channel it is matched with, but also the set of users matching with the same sub-channel. Similarly, each sub-channel not only considers which individual users to match with, but also that the subset of users has inner-relationship through power domain multiplexing. Thus, this is a many-to-many matching game with externalities, also known as peer effects~\cite{ECABA-2011}. $\hfill\blacksquare$

\section{}
\emph{Proof of Remark 2:} To be specific, given a sub-channel ${SC}_k \in \cal{K}$, for its most preferred user set ${\cal{S}}_k \subseteq \cal{M}$ that contains user $M_j$ and $M_i$ with ${\left| {{h_{k,i}}} \right|^2}/{n_{k,i}} > {\left| {{h_{k,j}}} \right|^2}/{n_{k,j}}$, if $M_i \in {\cal{S}}_k$, then it is not necessary that ${M_j} \in {\mathcal{S}_k}\backslash \left\{ {{M_i}} \right\}$. Due to the interference item, the value of $R_{k, j}$ may have changed after $M_i$ is removed from $\mathcal{S}_k$, and thus, ${SC}_k$ may not prefer $M_j$ any more, i.e., $\mathcal{S}_k$ is not necessary to remain the same. $\hfill\blacksquare$

\vspace{-0.2cm}

\section{}
\emph{Proof of Corollary 1:} If user $M_i$ and $M_j$ propose to switch their matches, then it implies that ${R_{p,i}}\left( \Psi  \right) \le {R_{q,i}}\left( {\Psi _{jq}^{ip}} \right)$ and ${R_{q,j}}\left( \Psi  \right) \le {R_{p,j}}\left( {\Psi _{jq}^{ip}} \right)$, in which at least one of equality does not hold. Note that any user ${M_k} \in {\mathcal{S}_p}\backslash \left\{ {{M_i}} \right\}$ can cancel both $M_i$'s and $M_j$'s message from its received signals over ${SC}_p$ since the channel gains of $M_i$ and $M_j$ are smaller than that of $M_k$. Thus, we have ${R_{p,k}}\left( \Psi  \right) = {R_{p,k}}\left( {\Psi _{jq}^{ip}} \right),\forall k \in {\mathcal{S}_p}\backslash \left\{ {{M_i}} \right\}$. Similarly, we have ${R_{q,t}}\left( \Psi  \right) = {R_{q,t}}\left( {\Psi _{jq}^{ip}} \right),\forall t \in {\mathcal{S}_q}\backslash \left\{ {{M_j}} \right\}$. Then we can derive the following inequality:
\begin{equation}
\begin{aligned}
{R_{S{C_p}}}\left( {\Psi _{jq}^{ip}} \right) & = \sum\limits_{k \in {\mathcal{S}_p}\backslash \left\{ i \right\}} {{R_{p,k}}} \left( {\Psi _{jq}^{ip}} \right) + {R_{p,j}}\left( {\Psi _{jq}^{ip}} \right) \\
& \ge \sum\limits_{k \in {\mathcal{S}_p}\backslash \left\{ i \right\}} {{R_{p,k}}} \left( \Psi  \right) + {R_{q,j}}\left( \Psi  \right) \\
&  = {R_{S{C_p}}}\left( \Psi  \right).
\end{aligned}
\end{equation}
Similarly, we have ${R_{S{C_q}}}\left( {\Psi _{jq}^{ip}} \right) \ge {R_{S{C_q}}}\left( \Psi  \right)$. Therefore, the swap matching ${\Psi _{jq}^{ip}}$ is approved by both the users and sub-channels. $\hfill\blacksquare$

\end{appendices}

\end{document}